\documentstyle[12pt]{article}

\begin{document}

\begin{center}
{\large {\bf Representations and BPS states of 10+2 superalgebra}}

\vspace{1cm} {\bf R. Manvelyan \footnote{{\bf E-mail: manvel@moon.yerphi.am}}
, A. Melikyan \footnote{{\bf E-mail: arsen@moon.yerphi.am}}, R. Mkrtchyan%
\footnote{{\bf E-mail: mrl@amsun.yerphi.am}}} \vspace{1cm} 

{\it Theoretical Physics Department,}

{\it Yerevan Physics Institute}

{\it Alikhanian Br. st.2, Yerevan, 375036 Armenia }
\end{center}

\vspace{1cm}

\begin{abstract}
The 12d supersymmetry algebra is considered, and classification of BPS
states for some canonical form of second-rank central charge is given. It is
shown, that possible fractions of survived supersymmetry can be 1/16, 1/8,
3/16, 1/4, 5/16 and 1/2, the values 3/8, 7/16 cannot be achieved in this
way. The consideration of a special case of non-zero sixth-rank tensor
charge also is included.
\end{abstract}
\vfill
{\smallskip \pagebreak }

\section{Introduction.}

One direction of recent development of supersymmetric theories in higher
dimensions is the consideration of 12-dimensional theories with signature
10+2 .$^{1-10}$ This dimension is the highest possible in a sense that
corresponding Lorentz group permits Mayorana-Weyl spinor $Q$, with 32 real
components, which upon reduction to 4d gives 8 spinors, appropriate to N=8
bound in four dimensions. Corresponding 12d susy algebra has the following
key relation

\begin{eqnarray}
\left\{ \overline{Q},Q\right\} &=&\Gamma ^{MN}Z_{MN}+\Gamma
^{MNPQRL}Z_{MNPQRL}^{+}  \label{qq} \\
M,N... &=&0,0^{^{\prime }},1,...10
\end{eqnarray}

which is of now common type, i.e. includes tensorial ``central'' charges, in
the second one superscript (+) means the self-dual part of tensor. The
non-usual feature is the absence of $\Gamma ^{M}P_{M}$ term, which is not
permitted by symmetry considerations. Algebra (\ref{qq}) may be considered
as, rewritten in condensed notations, the 11d M-theory superalgebra, in a
sense that changing a notation in a way appropriate for dimensional
reduction by second time dimension to 11d: 
\begin{equation}
Z_{\mu 0^{^{\prime }}}\Rightarrow P_{\mu },Z_{\mu \nu \rho \sigma
l0^{^{\prime }}}\Rightarrow Z_{\mu \nu \rho \sigma l}  \label{11d}
\end{equation}
we obtain the M-theory superalgebra .$^{11}$

\begin{equation}
\left\{ \overline{Q},Q\right\} =\Gamma ^{\mu }P_{\mu }+\Gamma ^{\mu \nu
}Z_{\mu \upsilon }+\Gamma ^{\mu \nu \lambda \rho \sigma }Z_{\mu \nu \lambda
\rho \sigma }  \label{qq11}
\end{equation}

So, any statement concerning (\ref{qq}) has a counterpart for (\ref{qq11}).
The 12d susy theories have been discussed in $^{1-9}$ where different aspects
are revealed, such as a connection to F-theory, reduction to 11d, etc.

The important problem is the construction of representations of 12d algebra (%
\ref{qq}). This problem actually reduces to the problem of finding the rank
of the matrix in the r.h.s at different values of charges $Z_{MN}$ and $%
Z_{MNPQRL}^{+}$. The diagonalization of that matrix brings (\ref{qq}) to the
form of an algebra of $n$ pairs of fermionic creation-annihilation
operators, where $n$ is half of the rank of matrix.

The main aim of the present paper is the construction and analysis of some
of the representations of algebra (\ref{qq}) in the case, mainly, when
sixth-rank tensor is equal to zero, and even in that case it is not possible
to obtain a complete answer, but we study some important cases. The problem
lie in obtaining a canonical form of antisymmetric tensor in (2+10)
dimensions. In addition, some cases with non-zero sixth rank charges are
analyzed.

When results for (\ref{qq}) are interpreted for (\ref{qq11}), we
obtain what is called on the language of M-brane an intersection of branes,
with some amount of supersymmetry maintained (and remaining part broken).
When some part of supersymmetry is maintained, the multiplet is called
shortened and states are called BPS. Many such states are presented by
classical solutions of corresponding supergravities equations, and possibly
all of them have such a representation, so our analysis shows  prospects
for the search of such  solutions.

Let's stress that such representation is possible only for 11d form 
(interpretation) of
algebra (1). For 12d case it is not possible, since the very existence of full
12d theory is not established up to now.

A number of results on the problem of calculating of the rank of of the
matrix in (\ref{qq}) are obtained in .$^{3,5,10,11}$

\section{Membrane}

In this section we consider the algebra (\ref{qq}) with non-zero membrane
charge only. The analysis is convenient to carry on in a language of
determinant of the matrix on r.h.s. of (\ref{qq}). More exactly, multiplying
(\ref{qq}) by $\Gamma ^{00^{^{\prime }}}$, we bring the problem to the
calculation of determinant of the matrix in the r.h.s. of 
\begin{equation}
\left\{ Q,Q\right\} =\Gamma ^{00^{^{\prime }}}\Gamma ^{MN}Z_{MN}  \label{qq3}
\end{equation}
Next step will be to bring the matrix $Z_{MN}$ to canonical form. In
euclidean space-time the canonical form would be the well-known form with
2x2 antisymmetric blocks on a main diagonal with eigenvalues $\lambda _{1}$ $%
\lambda _{2}$ $...$ $\lambda _{6}$. In pseudoeuclidean space-time some
matrixes cannot be brought to that form, and classification of canonical
forms is more complicated. The complete classification, which is connected
to classification of the orbits of coadjoint representation of corresponding
Lorentz group, will be discussed elsewhere. Nevertheless, the mentioned
quazidiagonal form is one of the canonical forms, and analysis in that case
can be carried up to the end. In that case the matrix has the following
form: 
\begin{equation}
\det \Gamma ^{00^{^{\prime }}}\Gamma ^{MN}Z_{MN}=\det (\lambda _{1}+\lambda
_{2}\Gamma ^{00^{^{\prime }}12}+...+\lambda _{6}\Gamma ^{00^{^{\prime
}}9\natural })  \label{det1}
\end{equation}
(Notation  $\natural$ is for tenth $\Gamma$-matrix:  $\Gamma^{\mu}$ at $\mu$=10).
Gamma-matrixes in r.h.s of this relation are commuting with each other,
hence diagonalizable.. Each of them has eigenvalues $\pm 1$ only, since
square of them is 1, and number of +1 is equal to number of -1, since trace
of them is zero. So, determinant in (\ref{det1}) is equal to product of
linear combinations of $\lambda _{i}$ with coefficients $\pm 1$. It may be
proved, that each combination enters in that product twice and nothing else
is contained. The number of combinations is $2^{5}=32$, which is exactly
half of a power of determinant in units of $\lambda $. So: 
\begin{equation}
\det \Gamma ^{00^{^{\prime }}}\Gamma ^{MN}Z_{MN}=\Pi (\lambda _{1}\pm
\lambda _{2}\pm ....\pm \lambda _{6})^{2}  \label{det2}
\end{equation}

Actually, for discussion of consequences of (\ref{det2}) on a representation
of (\ref{qq3}), we need an answer for determinants of operator in r.h.s in
subspaces of chiral spinors. The answer is the following: the product in (%
\ref{det2}) can be divided into two classes, with even and odd numbers of
sign changes in (\ref{det2}), and products in each classes are equal to
determinants in chiral subspaces. 
\begin{equation}
\det (\Gamma ^{00^{^{\prime }}}\Gamma ^{MN}Z_{MN})_{\pm }=\Pi _{\pm
}(\lambda _{1}\pm \lambda _{2}\pm ....\pm \lambda _{6})^{2}  \label{det3}
\end{equation}

where $\Pi _{\pm }$ denotes the product of combinations with even and odd
numbers of negative signs, respectively. From now on we consider the chiral
determinant with even number of sign changes: 
\begin{equation}
\Pi _{+}(\lambda _{1}\pm \lambda _{2}\pm ....\pm \lambda _{6})^{2}
\label{det4}
\end{equation}

Before starting an analysis of representations of (\ref{qq3}), it is
important to take into account the positivity property of l.h.s of (\ref{qq3}%
), which means that all eigenvalues of r.h.s. have to be non-negative (when
some of them are zero, this is just BPS cases). The requirement of
non-negativeness of all eigenvalues in the product of (\ref{det4}) leads to
the following inequalities, which have to be satisfied by $\lambda
_{1},\lambda _{2},...,\lambda _{6}$: 
\begin{equation}
\lambda _{1}\geq \mid \lambda _{2}\mid +...+\mid \lambda _{6}\mid  \label{l1}
\end{equation}

in the case when even number of $\lambda _{2},...,\lambda _{6}$ is positive,
and 
\begin{equation}
\lambda _{1}\geq \mid \lambda _{2}\mid +...+\mid \lambda _{5}\mid -\mid
\lambda _{6}\mid  \label{l2}
\end{equation}

when odd number of $\lambda _{2},...,\lambda _{6}$ is positive, and for
definiteness we assume that $\lambda _{6}$ has a minimal absolute value
among $\lambda _{2},...,\lambda _{6}$. When one of the $\lambda $-s is zero,
both inequalities coincide.

Now we turn to the discussion of representations of supersymmetry algebra (%
\ref{qq3}), which, as mentioned above,is based  actually on the calculation
of a numbers of pairs of fermionic creation-annihilation operators, which,
in turn, is equal to the half of the rank of matrix in the r.h.s. of (\ref
{qq3}). The relevant determinant is \ref{det4} and is already prepared for
discussion of it's rank, since is expressed as a product of eigenvalues.

First, for  general set of $\lambda $ determinant is non-zero, and dimension of
representation is $2^{16}$. These are not BPS states, since each
supersymmetry generator acts non-trivially on some of them. Next, when one
eigenvalue in (\ref{det4}) is zero, rank is lowered by 2. Dimension of
representation in the case of two zeros is $2^{14}$. This is already a BPS
state, and number of conserved supersymmetries is 2, i.e. 1/16 of initial
number, which was 32. Without loss of generality, we can assume that zero
eigenvalue is the sum of $\lambda$-s with positive sign: 
\begin{equation}
\lambda _{1}+\lambda _{2}+...+\lambda _{6}=0  \label{01}
\end{equation}

Now, the analysis of possibilities of appearing new zeros in (\ref{det4})
can be greatly simplified by taking into account the inequalities (\ref{l1}%
), (\ref{l2}). First, the case when one of $\lambda $ ($\lambda _{6}$,
namely) is , gives, as a result of (\ref{l1}), (\ref{l2}), (\ref{01}),
that new zeros of determinant (\ref{det4}) can appear only when some other $%
\lambda $ are zero. When total number of zeros among $\lambda
_{2},...,\lambda _{6}$ is 1, 2, 3, 4, then, as can be easily established,
number of zeros in product (\ref{det4}) is (counting (\ref{01}) also),
respectively, 2, 4, 8, 16. (All five $\lambda _{2},...,\lambda _{6}$ cannot
be zero, since from (\ref{01}) it follows that $\lambda _{1}=0$, which means
that we are considering the vacuum state.) So, the fractions of survived
supersymmetries, from the BPS point of view, are 1/16, 1/8, 1/4, 1/2,
respectively.

When all $\lambda _{2},...,\lambda _{6}$ are non-zero, then, if (\ref{l1})
is applicable, then there is no new zero eigenvalues in (\ref{det4}), which
easily follows from (\ref{01}). When inequality (\ref{l2}) is applicable,
then the number of zeros depends on the number of $\lambda $ among $%
\lambda _{2},...,\lambda _{5}$, with absolute values, equal to that of $%
\lambda _{6}$. When that number is 0, 1, 2, 3, 4, then the number of zeros
in (\ref{det4}), not forgetting (\ref{01}), is 2, 4, 6, 8, 10, respectively.
Correspondingly, the fractions of survived supersymmetries are 1/16, 1/8,
3/16, 1/4, 5/16, respectively.

One of the conclusions of this analysis is that the same fractions of
survived supersymmetries can be achieved in two different ways. That is
possible for the cases of fractions 1/16, 1/8, 1/4. Other values - 3/16,
5/16 and 1/2 can be achieved only by one of two mechanisms, presented above.
The values 3/8, 7/16 cannot be obtained in these cases.

As mentioned in Introduction, these results can be interpreted from the
point of view of 11d M-theory superalgebra.

\section{Inclusion of 6-form charges}

The number of Lorentz-invariant combinations grows strongly, when considering
 the last term in (\ref{qq}), i.e. the five-brane charge.
Instead of six $\lambda $, now we have in addition 462 invariants. The
above analysis can be extended in a simple way to the case of some non-zero
6-form charges, among these 462. That is the case, when corresponding
gamma-matrixes of 6-form charges commute with matrixes of 2-form charges.
There is 20 such matrixes, with 10 different coefficients, due to the
self-duality constraint. These are all possible products of $\Gamma
^{00^{^{\prime }}},\Gamma ^{12},...,\Gamma ^{9\natural }$.

The answers for full determinant, and for chiral determinants are the
following.

Let's denote the values of tensor $Z_{MNPQRL}^{+}$ in a coordinate system
where $Z_{MN}$ is already diagonal (in a sense of previous Section) by $%
z_{1},z_{2},...,z_{10}.$ Exactly, $z_{1}=Z_{00^{^{\prime }}1234}^{+},$ $%
z_{2}=Z_{00^{^{\prime }}1256}^{+},...,$ $z_{10}=Z_{00^{^{\prime
}}789,10}^{+} $. Components, dual to these, are equal to them, due to
self-duality, all other components of $Z_{MNPQRL}^{+}$ are equal to zero.

The all matrixes in r.h.s. of the (\ref{qq}) again are simultaneously
diagonalizable, and signs in front of new $z$ elements are a factors of the
signs in front of corresponding $\lambda $ (correspondence is evident - $%
z_{1}$ corresponds to $\lambda _{1},\lambda _{2}$ and $\lambda _{3}$, etc.).
So we can immediately write down the answer for corresponding determinant: 
\begin{eqnarray}
\det \Gamma ^{00^{^{\prime }}}(\Gamma ^{MN}Z_{MN}+\Gamma
^{MNPQRL}Z_{MNPQRL}^{+}) &=&  \nonumber \\
\det (\lambda _{1}+\lambda _{2}\Gamma ^{00^{^{\prime }}12}+...+\lambda
_{6}\Gamma ^{00^{^{\prime }}9,10}+\Gamma ^{00^{^{\prime }}}\Gamma
^{MNPQRL}Z_{MNPQRL}^{+}) &=&  \nonumber \\
\Pi _{+}(\lambda _{1}\pm \lambda _{2}\pm ....\pm \lambda _{6}\pm z_{1}\pm
z_{2}\pm ...\pm z_{10})^{2} &&  \label{zz}
\end{eqnarray}

where the signs of $z$ are chosen according to rule mentioned above,
i.e., the sign of given $z$ is the product of the signs of three
corresponding $\lambda $ (one of which is always $\lambda _{1}$, so one of
these signs is +1), and the product is, as above, over all even sign changes
of $\lambda $.

The positivity requirement in this case leads to the conditions on the set
of $\lambda $ and $z$, that eigenvalues in (\ref{zz}) all have to be
non-negative. The cases with some eigenvalues equal to zero are, as above,
the BPS cases. The condition of positivity in this case includes 16
inequalities on a 15 variables $\lambda ,z$, and cannot be written in a
simple form.

\section{Acknowledgments}

This work was supported in part by the U.S. Civilian Research and
Development Foundation under Award \# 96-RP1-253 and by INTAS grants \newline
\# 96-538 and \# 93-1038 (ext) .

\end{document}